\documentclass[letterpaper, 10 pt, conference]{ieeeconf}
\IEEEoverridecommandlockouts
\usepackage{amsmath,amssymb,mathrsfs,amsfonts}
\usepackage{algorithmic}
\usepackage{graphicx}
\usepackage{textcomp}
\usepackage{xcolor}
\usepackage{comment}
\usepackage{theorem}
\usepackage{hyperref}

\usepackage{tikz}
\usetikzlibrary{backgrounds}
\usetikzlibrary{intersections}
\usepackage{pgfplots}

\usepackage{blindtext}

\newtheorem{theorem}{Theorem}
\newtheorem{lemma}{Lemma}

\input{commands.sty}

\def\BibTeX{{\rm B\kern-.05em{\sc i\kern-.025em b}\kern-.08em
    T\kern-.1667em\lower.7ex\hbox{E}\kern-.125emX}}
\begin{document}
\title{\huge \bf Incentive Designs for Learning Agents to Stabilize Coupled Exogenous Systems}
\author{
Jair Cert\'orio, \IEEEmembership{Student Member, IEEE}, Nuno C. Martins, \IEEEmembership{Senior Member, IEEE}, \\ Richard J. La and Murat Arcak, \IEEEmembership{Fellow, IEEE}.
\thanks{ 
The work of Cert\'orio and Martins was supported by the AFOSR grant FA95502310467, and the NSF grants 2135561 and 2139713. The work of Arcak was supported by the NSF grant CNS-2135791
}
\thanks{
Jair Cert\'orio, Nuno C. Martins, and Richard J. La are with the Department of Electrical and Computer Engineering and Institute for Systems Research, University
of Maryland, College Park, College Park, MD 20740 USA (e-mail: {certorio, nmartins, hyongla}@umd.edu).}
\thanks{
Murat Arcak is with the Department of Electrical Engineering and
Computer Science, University of California, Berkeley, CA 94720 USA
(e-mail: arcak@eecs.berkeley.edu).
}
}

\maketitle

\begin{abstract} 

We consider a large population of learning agents noncooperatively selecting strategies from a common set, influencing the dynamics of an exogenous system (ES) we seek to stabilize at a desired equilibrium. Our approach is to design a dynamic payoff mechanism capable of shaping the population's strategy profile, thus affecting the ES's state, by offering incentives for specific strategies within budget limits. Employing system-theoretic passivity concepts, we establish conditions under which a payoff mechanism can be systematically constructed to ensure the global asymptotic stability of the ES's equilibrium. In comparison to previous approaches originally studied in the context of the so-called epidemic population games, the method proposed here allows for more realistic epidemic models and other types of ESs, such as predator-prey dynamics. The stability of the equilibrium is established with the support of a Lyapunov function, which provides useful bounds on the transient states.
\end{abstract}

\section{Introduction}
\label{sec:introduction}
Systems whose behavior depends on the strategic choices of many agents can be studied through the lens of evolutionary game theory, in particular when considering large populations of nondescript agents that repeatedly revise their strategies. 
Examples of systems with such dependency on the aggregate choice of a large population include models of traffic congestion \cite{arcak_dissipativity_2021}, optimal power dispatch \cite{pantoja_population_2011}, distributed task allocation \cite{park_multi-robot_2021}, building temperature control \cite{obando_building_2014}, and epidemic mitigation \cite{martins_epidemic_2023}.

The coupling between the \textit{evolutionary dynamics}, which models the population’s strategic choices, and the %
\textit{exogenous system} (ES) dynamics, which captures the state of the system affected by the decisions of the population, makes the task of designing stabilizing policies challenging. This is especially true when we aim to design policies that not only improve the behavior of the system, but also provide performance guarantees that hold at any given time.

In our work, we generalize the design concept from \cite{martins_epidemic_2023} to a larger class of ESs which, when agents stop revising their strategies, have a Lyapunov function and satisfy some mild assumptions. We design incentives that guarantee the convergence of the population state to an equilibrium, which can be selected independently of the payoff mechanism. While our results require similar assumptions on the behavior of the population to those of \cite{martins_epidemic_2023}, unlike \cite{martins_epidemic_2023,certorio_epidemic_2022}
we obtain a bound on the instantaneous cost of implementing our incentives. 

In addition, we show that the proposed payoff mechanism is compatible with many ESs, such as the epidemic models studied by \cite{martins_epidemic_2023,certorio_epidemic_2022,certorio_epidemic_2023} and the Leslie-Gower model for studying the interaction between populations of hosts and parasites \cite{korobeinikov_lyapunov_2001}. 
As an application, we use the proposed framework to devise incentives for a modified epidemic model studied in \cite{certorio_epidemic_2022}, while considering disease transmission rates that depend nonlinearly on the agents' choices.

\subsection{Contributions}

The goal of our study is to develop a new framework for designing a dynamic payoff mechanism that guarantees the convergence of both the population and the ES, whose dynamics are influenced by the strategic choices of the agents in the population, to a desirable equilibrium that can be selected independently of the payoff.
The incentive we design has a bound on the instantaneous reward offered to the population, which is not guaranteed in the previous studies that use $\delta$-passivity for designing policies that mitigate epidemics. 

The assumptions we introduce on the learning rule employed by the agents are similar to those of \cite{martins_epidemic_2023}. However, we relax one of the assumptions on the learning rule, which is not easy to verify, and replace it with another assumption that is easier to check. 
The proposed mechanism works on previously studied ESs \cite{martins_epidemic_2023,certorio_epidemic_2022,certorio_epidemic_2023}, but is not limited to epidemic models; our framework is more general and can be applied to any system satisfying the conditions identified in this paper, even when the learning rule is unknown to the policy maker.
As an illustrative example, we show in \S \ref{ssec:Korobeinikov} that the Leslie-Gower system studied in \cite{korobeinikov_lyapunov_2001} satisfies these conditions.

\section{Related Works}
\label{sec:related_works}

Earlier studies showed that, for potential games, bounded-rational learning rules with the \textit{positive correlation} (PC) property, where agents revise their strategies in a way that increases their payoffs, guarantee the convergence of the population state to \textit{Nash equilibria} (NEs) \cite{sandholm_potential_2001}. In addition, Hofbauer and Sandholm \cite{hofbauer_stable_2009} established that, for the class of \textit{contractive games}, many evolutionary dynamics lead the population state to an NE. A recent study demonstrated that for certain potential and strictly contractive games,  the population state converges to an NE, even when the revision rates depend explicitly on the current strategies of the agents \cite{kara_population_2022,kara_excess_2023}.
For a survey of earlier studies and the applications of population games we refer the reader to \cite{sandholm_population_2010,quijano_role_2017}, and the references therein.

Motivated by the class of contractive games, Fox and Shamma \cite{fox_population_2013} showed that certain learning rules, such as \textit{impartial pairwise comparison} and \textit{excess-payoff target} rules, exhibit a form of passivity which they named $\delta$-passivity. The concept of $\delta$-passivity was generalized in  \cite{park_payoff_2020} to admit a large class of dynamical payoff mechanism
and in \cite{arcak_dissipativity_2021} that introduced $\delta$-dissipativity. 
In \cite{schweidel_compositional_2022}, the authors determined a sufficient condition for the interconnections of $\delta$-dissipative dynamical systems to also be $\delta$-dissipative.

Adapting tools from robust control, Mabrok and Shamma \cite{mabrok_passivity_2016} studied the passivity properties of higher-order games and determined necessary conditions for evolutionary dynamics to be stable for all higher-order passive games. Their work also proved that replicator dynamics is lossless \cite{mabrok_passivity_2016}, which was later shown to be not $\delta$-passive \cite{park_passivity_2018}.

The population game framework has been used in many problems. For example, it has been used for distributed optimization \cite{martinez-piazuelo_payoff_2022} and distributed NE seeking \cite{martinez-piazuelo_distributed_2022}.  %
Obando et al. \cite{obando_building_2014} 
studied a temperature control problem, where the population models the heating power to be distributed in a building and is coupled to a thermal model of the building. We refer the reader to a survey  \cite{quijano_role_2017} for additional examples.

To the best of our knowledge, \cite{martins_epidemic_2023} is the first study that used $\delta$-passivity as a design tool:
a dynamic payoff mechanism was designed to lessen the impact of an epidemic subject to a limit on the long-term budget available to the decision maker. 
This work was extended to cases with nonnegligible disease mortality rates \cite{certorio_epidemic_2022} and to scenarios with noisy payoffs to agents~\cite{park2024epidemic}. The same framework was also used to consider two-population scenarios~\cite{certorio_epidemic_2023}. 

Our work extends the design method in \cite{martins_epidemic_2023} to a larger class of ESs, which includes epidemic models as examples. Our assumptions are similar to those of previous studies, but the proposed dynamic payoff mechanism has a provable bound on the incentives provided to the agents. We also determine conditions under which a class of ESs coupled to a population of learning agents can be stabilized to a desired equilibrium.

\section{Population Games and Learning Rules}

We consider a population of a large number of nondescript agents, in which each agent follows a single strategy at any given time and can repeatedly revise its strategy, based on the payoffs to available strategies at the revision times. We assume that the agents have a common set of $n$ strategies available to them. The instantaneous payoff obtained by following the $i$-th strategy at time $t$ is given by $p_i\st$, and the payoff vector offered to the population at time $t$ is denoted as $p\st := (p_i\st \;|\; i \in [n])$, where $[n]:=\{1,\dots,n\}$. The payoff perceived by agents at time $t$ is the difference between the rewards offered by the policy maker at time $t$, which is denoted by $r\st$, and the vector that contains the intrinsic costs of the $n$ strategies, which we denote by $c$. Thus, the payoff vector at time $t$ is given by 
\begin{align}
    p\st = r\st-c.
\end{align}

These assumptions render the tools of population games well-suited for analyzing the strategic interactions among the agents. 
The population state at time $t$ is denoted by $x\st := (x_i\st \;|\; i \in [n])$, with $x_i\st$ being the proportion of the population following the $i$-th strategy at time $t$. The vector $x\st$ takes values in the standard simplex
\begin{align*}
    \mathbb{X} := \left\{ x \in [0,1]^n \; \Big|\; \sum_{i=1}^{n} x_i = 1 \right\}.
\end{align*}

In the large-population limit, for $t\geq0$, the population state $x$ evolves according to the Evolutionary Dynamics Model (EDM)
\begin{align} \tag{EDM}
    \dot{x}\st &= V(x\st,p\st),
\end{align}
where $V:\mathbb{X}\times \mathbb{R}^n \rightarrow \mathbb{R}^n$. The
$i$-th element of $V(x,p)$ is 
\begin{align}
    V_i(x,p) := \sum_{j=1}^n \left( x_j\tau_{ji}(x,p) -x_i \tau_{ij}(x,p)  \right), \label{eq:EDM_dynamics}
\end{align}
with a learning rule (also referred to as a revision protocol) $\tau$ that is a Lipschitz continuous map $\tau:\mathbb{X}\times\mathbb{R}^n \rightarrow [0,\bar{\tau}]^{n \times n}$, upper bounded by $\bar{\tau}>0$. 

We consider the EDM coupled to an ES,
whose state at time $t$ is denoted
by $y(t)$. The ES state $y(t)$ takes values in $\mathbb{Y} \subset \mathbb{R}^m$ and evolves according to
\begin{align}
    \dot{y}\st = f(y\st;x\st), \label{eq:system_f}
\end{align}
where $f: \mathbb{R}^m\times\mathbb{R}^n \rightarrow \mathbb{R}^m$ is locally Lipschitz continuous and the population state $x\st$ acts as time-varying parameters of the ES.  We assume that, for any $x \in \mathbb{X}$, if $x\st \equiv x$ then \eqref{eq:system_f} has a unique equilibrium denoted as $y^*(x)$.

The dynamics of rewards $r\st$ offered to the agents is described by the following:
\begin{align}
    \dot{q}\st &= G(y\st, x\st, q\st),          
                \label{eq:q_dynamics}\\
    r\st &= H(y\st, x\st, q\st), \nonumber \\
    q\sta{0} &= q_0, \nonumber
\end{align}
where $G$ and $H$ form a dynamic payoff mechanism to be designed by the policy maker.

\begin{definition}
    A learning rule $\tau$ is said to satisfy the \textit{positive correlation} (PC) condition if the following holds: for all $(x,p)\in\mathbb{X}\times\mathbb{R}^n$,
    \begin{align*}
        V(x,p)\neq 0 \;\Rightarrow\;  p^\top V(x,p) > 0.  
    \end{align*}
\end{definition}

\begin{definition}
    A learning rule $\tau$ is \textit{Nash Stationary} (NS) if, given the best response map $\mathscr{M}: \mathbb{R}^n \;\rightarrow\; 2^{\mathbb{X}}$, where 
    \begin{align*}
        \mathscr{M}(p) := \argmax_{x \in \mathbb{X}}\; p^\top x, \quad p \in \mathbb{R}^n,
    \end{align*}
    the following holds:
    \begin{align*}
        V(x,p)=0 \;\Leftrightarrow\;  x \in \mathscr{M}(p),\; p \in \mathbb{R}^n.
    \end{align*}
\end{definition}

\begin{definition}
    An EDM is $\delta$-passive if there exist (i) a differentiable function
    $\mathcal{S}:\mathbb{X}\times\mathbb{R}^n \;\rightarrow\; \mathbb{R}_{\geq0} $ and (ii) a Lipschitz continuous function
$\mathcal{P}:\mathbb{X}\times\mathbb{R}^n \;\rightarrow\; \mathbb{R}_{\geq0}$, which satisfy the following inequality for
all $x$, $p$ and $u$ in $\mathbb{X}$, $\mathbb{R}^n$ and $\mathbb{R}^n$, respectively:
\begin{align} \label{eq:storageIneq}
    \hspace{-0.1in} 
    \frac{\partial \mathcal{S}}{\partial x}(x,p) V(x,p)+ \frac{\partial \mathcal{S}}{\partial p}(x,p) u\leq - \mathcal{P}(x,p) +u^\top V(x,p) 
\end{align}
where $\mathcal{S}$ and $\mathcal{P}$ must also satisfy the equivalences below:
\begin{align*}
    \mathcal{S}(x,p)=0 \;\Leftrightarrow\; V(x,p)=0,\\
    \mathcal{P}(x,p)=0 \;\Leftrightarrow\; V(x,p)=0.
\end{align*}    
\end{definition}

Since the EDM is determined by the learning rule, we say that the learning rule is $\delta$-passive if the resulting EDM is $\delta$-passive.
 Two well-known classes of learning rules that satisfy PC and NS conditions and lead to $\delta$-passive evolutionary dynamics are the \textit{separable excess payoff target} and the \textit{impartial pairwise comparison} learning rules.

\begin{example} A learning rule $\tau$ is said to be of the {separable excess payoff target} type \cite{sandholm_excess_2005} if, for each $j$ in $[n]$, there is some $\rho_j:\mathbb{R}\rightarrow \mathbb{R}_{\geq0}$ such that
\begin{align*}
    \tau_{ij}(x,p) &= \rho_j(p_j-x^\top p) \ \mbox{ for all } i \in [n], 
\end{align*}
and $\rho_j$ satisfies $\rho_j(v)=0$ for $v\leq0$ and $\rho_j(v)>0$ for $v>0$.
\end{example}
\begin{example} A learning rule is said to be of the {\em impartial pairwise comparison} type \cite{sandholm_pairwise_2010} if, for each $j$ in $[n]$, there is some $\rho:\mathbb{R}\rightarrow \mathbb{R}_{\geq0}$ such that
\begin{align*}
    \tau_{ij}(x,p) &= \rho_j(p_j-p_i) \ \mbox{ for all }
    i \in [n], 
\end{align*}
and $\rho_j$ satisfies $\rho_j(v)=0$ for $v\leq0$ and $\rho_j(v)>0$ for $v>0$.
\end{example}

Lastly, we introduce a lemma that will be useful for proving our main result (Theorem~\ref{thm:convergence}) in the following section.

\begin{lemma}
    \label{lemma:uniqueBestResponse}
    For any fixed $q \in \mathbb{R}^n$ and $\bar{x} \in \mathscr{M}\left(q\right)$, the only vector $x \in \mathbb{X}$ that satisfies 
    \begin{align}
        x \in \mathscr{M}\left(\bar{x}-x+q\right) \label{eq:uniqueBestResponse}
    \end{align}
    is $x = \bar{x}$.
\end{lemma}
\begin{proof}
Since $\bar{x} \in \mathscr{M}\left(q\right)$, it is a solution to (\ref{eq:uniqueBestResponse}). To see that no other solution exists, rewrite
(\ref{eq:uniqueBestResponse}) as
$$
(y-x)^\top \left(x-\bar{x}-q\right) \ge 0 \quad \forall y\in \mathbb{X}.
$$
Define $f(x) = \|x-\bar{x}-q\|_2^2$, and note from the inequality above 
$$
(y-x)^\top \nabla f(x)\ge 0 \quad \forall y\in \mathbb{X}.
$$
It then follows from the minimum principle that
$x\in \argmin f(z)$ s.t. $z\in \mathbb{X}$. As this is a convex problem with strictly convex objective, it has a unique solution. Since $\bar{x}$ is a solution as noted above, the lemma follows.
\end{proof}

\section{Main Result}

Our goal is to design a dynamic payoff mechanism given by the maps $G$ and $H$, which not only guarantees the convergence of the population state to some $x^* \in \mathbb{X}$ selected by the policy maker, but also gives bounds on the ES state $y\st$ and the instantaneous cost $r\st$ incurred by the policy maker. 
Our convergence result assumes that the population adopts a learning rule that is $\delta$-passive, NS and PC, along with some conditions on the ES in~\eqref{eq:system_f}. %
For the incentive design, we do not need to know the learning rule used by the agents as long as it satisfies the properties above.

Although our assumptions are similar to those in the previous works, we do not require the assumption in equation \cite[(13)]{martins_epidemic_2023} on the storage function associated with the learning rule, but instead assume that the learning rule is PC, which is simpler to check.

Our approach to designing the dynamic payoff mechanism leverages the $\delta$-passivity of the EDM, which yields a Lyapunov function for the overall system. The Lyapunov function is used to bound $(y,x)\st$ based on the initial condition $(y,x)\sta{0}$. Moreover, the maps $G$ and $H$, combined with the bound on $(y,x)\st$, also enable us to bound the instantaneous rewards provided to the agents, as discussed in \S \ref{sec:bounds}.

\begin{theorem} \label{thm:convergence}
Consider a payoff vector $p^*\in\mathbb{R}^n$, a population state $x^* \in \mathscr{M}(p^*)$, and positive design parameters $k_1, k_2,$ and $k_3$. Suppose that the exogenous system \eqref{eq:system_f} satisfies the following:
    
    (i) There is a nonnegative continuously differentiable function $\mathcal{U}:\mathbb{Y} \times \mathbb{X} \rightarrow \mathbb{R}_{\geq0}$ such that, for any $x \in \mathbb{X}$ and $y \in \mathbb{Y}$, $\frac{\partial \mathcal{U}}{\partial y}(y;x) f(y;x)\leq 0$.
    
    (ii) For any $\alpha \in \mathbb{R}_{\geq 0}$, $\{ (y,x) \in \mathbb{Y} \times \mathbb{X} \; | \; \mathcal{U}(y;x)\leq \alpha \}$ is compact.

    (iii) For every $x \in \mathbb{X}$, the set $\{ y^*(x) \}$ is the largest invariant subset of $\{y \in \mathbb{Y} | \frac{\partial \mathcal{U}}{\partial y}(y; x)f(y;x) = 0\}$.  
        
    (iv) The function $\mathcal{U}$ satisfies $\frac{\partial \mathcal{U}}{\partial x}(y^*(x);x)=\mathbf{0}$ for every $x \in \mathbb{X}$.

    \noindent In addition, assume that 
    
    (v) the learning rule $\tau$ is Nash stationary, $\delta$-passive, and positively correlated.

    \noindent Then, the dynamic payoff mechanism given by 
    \begin{align}
        G(y, x, q) = & -k_1\nabla_x \mathcal{U}(y;x) \nonumber \\ 
        & -k_2(x-x^*) -k_3 (q-p^*), \label{eq:DPM} \\
        H(y, x, q) &= c + q, \nonumber
    \end{align}
    guarantees that, for any initial condition $(y,x)\sta{0} \in \mathbb{Y}\times\mathbb{X}$ and $q_0$ in $\mathbb{R}^n$, 
    we have $(y,x,q)\st \xrightarrow{t \rightarrow \infty } (y^*(x^*),x^*,p^*)$.
\end{theorem}
\begin{proof}
    Define the following candidate Lyapunov function for the overall system comprised of \eqref{eq:EDM_dynamics}, \eqref{eq:system_f}, and \eqref{eq:q_dynamics}. 
    \begin{align}
        \mathcal{L}(y,x,q) :=& \ k_1\mathcal{U}(y;x) + k_3(\max_i(p_i^*)- x^\top p^*) \nonumber \\ & +\frac{k_2}{2} \|x-x^*\|_2^2 + \mathcal{S}(x,q), \label{eq:Lyapunov_Fun}
    \end{align}
    where $\mathcal{S}$ is the storage function of the EDM. Due to our selection of $H$, we have $p=q.$ 

    We denote by $\dot{\mathcal{L}}(y,x,q)$ the directional derivatives of the function \eqref{eq:Lyapunov_Fun} along the vector field defined by \eqref{eq:EDM_dynamics}, \eqref{eq:system_f}, and \eqref{eq:q_dynamics} at the point $(y,x,q)$:
    \begin{align*}
        \dot{\mathcal{L}}(y,x,q) 
        = & k_1 (\nabla_y \mathcal{U}(y; x)^\top f(y;x)+ \nabla_x \mathcal{U}(y; x)^\top V(x,q)) \\ & - k_3 V(x,q)^\top p^* + k_2(x-x^*)^\top V(x,q) 
        \\ & + \nabla_x \mathcal{S}(x,q)^\top f(y;x) + \nabla_p \mathcal{S}(x,q)^\top  G(y, x, q)\\
        \leq &  k_1\nabla_y \mathcal{U}(y; x)^\top f(y;x) + V(x,q)^\top G(y,x,q) \\  &+ (k_1 \nabla_x \mathcal{U}(y; x)+k_2(x-x^*) -k_3 p^*)^\top V(x,q) \\ &- \mathcal{P}(x,q) \nonumber, \\
        = & k_1 \nabla_y \mathcal{U}(y; x)^\top  f(y;x) - k_3 q^\top V(x,q) - \mathcal{P}(x,q),
    \end{align*}
    where the inequality follows from \eqref{eq:storageIneq}. 
    Conditions (i) and (v) imply that $\dot{\mathcal{L}}$ is nonpositive, %
 and $\mathcal{L}$ is a nonstrict Lyapunov function.

    By condition (ii) and $\dot{\mathcal{L}} \leq 0$, $\{(y,x)\st | t \geq 0\}$ is bounded. Since \eqref{eq:DPM} is a bounded-input bounded-output linear system with state $p$ and bounded input $-k_1\nabla_x \mathcal{U}(y;x)-k_2(x-x^*)+k_3 p^*$, which is a continuous function of the trajectory $\{(y,x)\st | t \geq 0\}$, we obtain that $\{(y,x,q)\st | t \geq 0\}$ is bounded.

    Let $\mathbb{E}$ denote the largest invariant subset within $\{(y,x,q) \;|\; \dot{\mathcal{L}}(y,x,q) = 0\}$.
    From the LaSalle-Krasovskii invariance principle, $(y,x,q)\st$ converges to the $\omega$-limit set $L^+ \subset \mathbb{E}$, which is compact and invariant with respect to \eqref{eq:EDM_dynamics}, \eqref{eq:system_f}, and \eqref{eq:q_dynamics}  \cite[Lemma 4.1]{khalil_nonlinear_1995}.

    For any $(y,x,q)\sta{0} \in \mathbb{E}$, the trajectory will satisfy that, for all $t\geq0$, $x\st=x\sta{0}$ and $y\st=y^*(x\sta{0})$ due to the fact that $\dot{\mathcal{L}}(y\st,x\st,q\st) = 0$ for all $t\geq 0$ implies $\dot{x}\st=0$ for all $t\geq0$, and condition (iii) in Theorem~\ref{thm:convergence}. This, together with condition (iv), leads to
    \begin{align*}
        \dot{q}\st = -k_2(x\sta{0}-x^*)- k_3 (q\st-p^*)
    \end{align*}
    so that $q\st \xrightarrow{t \rightarrow \infty}\frac{k_2}{k_3}(x^*-x\sta{0})+p^*$ and  
    \begin{align*}\textstyle
        (y,x,q)(t) \xrightarrow{t \rightarrow \infty} (y^*(x\sta{0}),x\sta{0},\frac{k_2}{k_3}(x^*-x\sta{0})+p^*).
    \end{align*}
    
    Since the learning rule is assumed NS, we must have 
    \begin{align*}\textstyle
        x\sta{0} \in \mathscr{M}\left(x^*-x\sta{0}+\frac{k_3}{k_2}p^*\right),
    \end{align*}
    and by Lemma \ref{lemma:uniqueBestResponse} we have $x\sta{0}=x^*$ and $L^+ = \{(y^*(x^*),x^*,p^*)\}$. 
\end{proof}

\section{Bounds}
\label{sec:bounds}

As proven in Theorem~\ref{thm:convergence}, if the ES in~\eqref{eq:system_f} satisfies (i)-(iv) and the learning rule %
 satisfies (v), the payoff mechanism described in \eqref{eq:DPM} can be used to stabilize the ES and the population to a desired equilibrium. Also, we are able to determine bounds for the state and the rewards $r\st$ offered to the population.

 If $p\st=\mathbf{0}$ at some $t$, the storage function of the EDM is equal to zero at $t$, as any $x\st \in \mathbb{X}$ is a best response. 
 In particular, %
 if $p(0) = p^* = \mathbf{0}$, both $\mathcal{S}(x\sta{0},p\sta{0})$ and $\max(p^*)-x\sta{0}^\top p^*$  are equal to zero, and we have  
\begin{align*}
    \mathcal{L}_0 &:= \mathcal{L}(y\sta{0},x\sta{0},\mathbf{0}) \\
    &= k_1\mathcal{U}(y\sta{0};x\sta{0}) + \frac{k_2}{2} \|x\sta{0}-x^*\|_2^2.
\end{align*}
Furthermore, because $\mathcal{L}$ is decreasing along trajectories and $\mathcal{S}(x,\mathbf{0})=0$ for any $x \in \mathbb{X}$, we obtain that, for any $t\geq0$,
\begin{align}
    \mathcal{L}(y\st,x\st,\mathbf{0}) \leq \mathcal{L}(y\st,x\st,p\st) &\leq \mathcal{L}(y\sta{0},x\sta{0},\mathbf{0}).  \label{ineq:L_less_L0}
\end{align}
This in turn can be used to bound not only $y\st$ and $x\st$ but also the policy maker's instantaneous cost $\Bar{c}\st:= x\st^\top r\st$: for any $t\geq0$,
\begin{align}
    & \Bar{c}\st \leq \max\{ g(y,x)  | x\in \mathbb{X}, y\in \mathbb{Y}, \mathcal{L}(y,x,\mathbf{0})\leq \mathcal{L}_0 \}, \label{ineq:p_bound} \\
    & g(y,x) := \left\|  c+G(y, x, \mathbf{0})/k_3\right\|_\infty, \nonumber
\end{align}
where the terms $G(y, x, \mathbf{0})$, $\mathcal{L}(y,x,\mathbf{0})$ and $\mathcal{L}_0$ are affected by the choice of the parameters $k_1$, $k_2$, and $k_3$.

\section{Examples}

In \S\ref{ssec:Korobeinikov} and \S\ref{ssec:epg} we present two systems that fit our framework as the ES in \eqref{eq:system_f}.  They meet conditions (i)-(iv) of Theorem \ref{thm:convergence}, and when coupled to a population that employs a learning rule satisfying condition (v), the dynamic payoff mechanism described in Theorem \ref{thm:convergence} stabilizes the overall system to a desired equilibrium $y^*(x)$. In \S\ref{ssec:application} we consider a modification of \cite{certorio_epidemic_2022} to exemplify how our theorem can be leveraged for design: We first select a target equilibrium population state $x^*$ that minimizes the disease transmission rate subject to a budget constraint, and choose the parameters $k_1, k_2$, and $k_3$ so that the peak size of the infected population %
is guaranteed to be below a given threshold. We then present simulation results using several different learning rules.

Our examples focus on systems that are naturally coupled to a population of agents and are affected by the strategic choices of the agents. The system considered in \cite{martins_epidemic_2023,certorio_epidemic_2022} is a compartmental model of an epidemic disease, and the population state affects the transmission rate of the disease. Similarly, \cite{certorio_epidemic_2023} considers an epidemic model with two interacting populations, with the transmission rates of each population being affected by its agents' current strategies. 
Korobeinikov \cite{korobeinikov_lyapunov_2001} studied a Host-Parasite model, and by finding a nonstrict Lyapunov function he proved the convergence to the unique equilibrium of the model. We modify this model so that some of its parameters change according to the population state. We choose a desirable equilibrium $(y^*(x^*),x^*)$ of this modified model to reduce the number of parasites at the equilibrium, and then use \eqref{eq:DPM} to stabilize the equilibrium.

\subsection{Leslie-Gower predator-prey model}
\label{ssec:Korobeinikov} 
Korobeinikov \cite{korobeinikov_lyapunov_2001} studies the Leslie-Gower model that captures the interaction of populations of hosts and parasites. 
Let $O\st$ and $P\st$ denote the number of hosts and parasites, respectively, at time $t$. The population sizes evolve according to the following differential equations:
\begin{subequations} \label{eq:HP}
    \begin{align}
        \dot{O}\st &= (z_1 -a_1 P\st - b_1 O\st)O\st, \\
        \dot{P}\st &= (z_2 -a_2 P\st/O\st )P\st,
    \end{align}
\end{subequations} 
where $z_1,z_2,a_1$, and $a_2$ are positive, and $b_1$ is nonnegative. 
The intrinsic population growth rates of the hosts and parasites are $z_1$ and $z_2$, respectively.
The parameter $b_1$ relates to a growth limit on the hosts without parasites, while $a_1$ relates to a decrease of hosts due the parasites, and 
$a_2$ relates to a population limit on the parasites due to the number of  hosts.  
The unique co-existing equilibrium, where $O^*,P^*>0$, is
    \begin{align*}
        O^* = \frac{z_1}{a_1 z_2 + a_2 b_1} \ \mbox{ and } \
        P^* = \frac{z_1 z_2}{a_1 z_2 + a_2 b_1}.
    \end{align*}

The following is a nonstrict Lyapunov function of \eqref{eq:HP} on $(0,\infty)^2$: 
\begin{align}
    U(O,P) := \tilde{\mathcal{U}}((O,P);(O^*,P^*)), \label{eq:HP_Lyap}
\end{align}
where
\begin{align*}
    \tilde{\mathcal{U}}(v;w) := \log\left(\frac{v_1}{w_1}\right) + \frac{w_1}{v_1} + \frac{a_1 w_1}{a_2} \left(\log\left(\frac{v_2}{w_2}\right)+\frac{w_2}{v_2}\right),
\end{align*}
with $v,w \in \mathbb{R}_{>0}^2$. The directional derivatives of \eqref{eq:HP_Lyap} along the vector field defined by \eqref{eq:HP} is 
\begin{align*}
    \dot{U}(O,P) = - \frac{a_1}{P}(P-P^*)^2 - \frac{b_1}{O}(O-O^*)^2 %
\end{align*}
with positive $a_1$ and nonnegative $b_1$. This confirms that \eqref{eq:HP_Lyap} is a Lyapunov function of \eqref{eq:HP}.

Suppose that $z_1$ and $z_2$ are functions of the population state, i.e., $z_1, z_2:\mathbb{X}\rightarrow \mathbb{R}_{>0}$, and $b_1: \mathbb{X}\rightarrow \mathbb{R}_{\geq0}$.  Such scenarios could arise when the agents are the farmers who breed and raise livestock, which are the hosts affected by parasites. The strategic choices of the agents could include, for example, how many animals to breed or which measures to take to reduce the spread of parasites, e.g., diagnosing, isolating, and treating infected hosts. 
In this case, the Leslie-Gower model as the ES in \eqref{eq:system_f} is described by
\begin{align*} %
    f(y;x) :=& \begin{bmatrix}
        (z_1(x) -a_1 y_2 - b_1(x) y_1)y_1\\
        (z_2(x) -a_2 y_2/y_1 )y_2
    \end{bmatrix}
\end{align*}
with the equilibrium
\begin{align*}
    y^*(x) :=& \begin{bmatrix}
    \frac{z_1(x)}{a_1 z_2(x) + a_2 b_1(x)} \\
    \frac{z_1(x) z_2(x)}{a_1 z_2(x) + a_2 b_1(x)}
    \end{bmatrix} \in \mathbb{Y}=(0,\infty)^2.
\end{align*} 
Suppose $z_1, z_2$, and $b_1$ are continuously differentiable so that the equilibrium map $y^*$ is also continuously differentiable.

Define
\begin{align*}
    \mathcal{U}(y;x) := \tilde{\mathcal{U}}(y;y^*(x))-\frac{a_1}{a_2}y_1^*(x) \ .
\end{align*} 
Note that (a) $\mathcal{U}(y;x)>0$ for any $y \neq y^*(x)$, and (b) for any $x \in \mathbb{X}$, if $x\st \equiv x$, then $\dot{\mathcal{U}}(y\st;x\st)=\dot{U}(y_1\st,y_2\st)$. Thus, it satisfies conditions (i) and (iii). %
As a continuous function, the sublevel sets of $\mathcal{U}$ are closed, and we can verify that they are also bounded so that condition (ii) is satisfied. 
Lastly, $\mathcal{U}$ satisfies (iv):
\begin{align*}
    &\frac{\partial \mathcal{U}}{\partial x}(y^*(x);x) \\ 
    & = \left(\frac{\partial \tilde{\mathcal{U}}}{\partial w}(y^*(x);y^*(x)) -\frac{a_1}{a_2} \begin{bmatrix}
        1\\0
    \end{bmatrix}^\top\right)  \frac{\partial y^*}{\partial x}(x)=\mathbf{0},
\end{align*}
because
\begin{align*}
    \frac{\partial \tilde{\mathcal{U}}}{\partial w}(y^*(x);y^*(x)) = \frac{a_1}{a_2} \begin{bmatrix}
        1\\0
    \end{bmatrix}^\top.
\end{align*}

\subsection{Epidemic Population Games (EPG)}
\label{ssec:epg}

Previous studies on EPG \cite{martins_epidemic_2023,certorio_epidemic_2022,certorio_epidemic_2023} examined epidemic compartmental models coupled to a population, with the epidemic model being the ES. Here we show that the model used in \cite{certorio_epidemic_2022} satisfies the conditions in Theorem~\ref{thm:convergence}, even for a more general dependency of the transmission rates on the agents' strategies than that considered in \cite{certorio_epidemic_2022}. The epidemic model satisfies conditions (i)-(iv) and, when coupled to a population employing a learning rule that satisfies (v), we can use Theorem \ref{thm:convergence} to drive them to a desirable equilibrium. A similar analysis shows that the epidemic models in  \cite{martins_epidemic_2023,certorio_epidemic_2023} also satisfy conditions (i)-(iv).

We first briefly describe the normalized susceptible-infectious-recovered-susceptible (SIRS) model, which is the ES we aim to stabilize. Let $I\st$, and $R\st$ denote the proportions of infected agents and recovered agents, respectively, in the population at time $t$. Suppose that $N\st$ is the population size at time $t$. The population size changes according to $\dot{N}\st = (g-\delta I\st)N\st$, where $g:= \theta - \zeta$ is the difference between the birth rate $\theta$ and the natural death rate $\zeta$, and $\delta>0$ is the disease death rate. The disease recovery rate and the rate at which a recovered individual becomes susceptible again due to waning immunity are denoted by $\gamma$ and $\psi$, respectively. 

Since the model is normalized, $I\st \in [0,1]$ and $R\st \in [0,1-I\st]$ at any time $t$. The ES state $y\st$ is given by $(I\st,R\st)$ and evolves according to
\begin{subequations} \label{eq:SIRS}
\begin{align}
    \dot{I}\st &= (\cB\st S\st  +\delta I\st - \sigma) I\st, \label{eq:SIRSa} \\ \label{eq:SIRSb}
    \dot{R}\st &= \gamma I\st - \omega R\st +\delta R\st I\st,
\end{align} 
\end{subequations} 
where $\cB\st$ is the average transmission rate at time $t$,
$\bar{\sigma}:=\gamma+\zeta+\delta$, $\sigma:=g+\bar{\sigma}$, $\bar{\omega}:=\psi+\zeta$, $\sigma:=g+\bar{\omega}$, and $\bar{\sigma}^{-1}$ is the mean infectious period for an affected individual (till recovery or death). 
The adopted time unit is one day, and newborns are assumed susceptible. As in \cite{certorio_epidemic_2022}, we assume $\delta>0$ but moderate such that $\delta < \min \{ \omega, \gamma \}$. Also, $\cB\st>\sigma$ for all $t\geq 0$ so that there is a unique endemic equilibrium.

For fixed $\cB\st \equiv B > \sigma$, the endemic equilibrium of \eqref{eq:SIRS} is given by the differentiable functions of $B$:
\begin{subequations} 
\begin{align*}
    I_B^* :=&  \frac{ b_B - \sqrt{\Delta}}{2\delta (B-\delta)},  \ \mbox{ and}\\
    R_B^* :=& (1 - \sigma / B ) - (1 - \delta /  B ) I_B^*,
\end{align*}
\end{subequations} 
where $b_B := \gamma B + \omega (B - \delta) + \delta(B - \sigma)$, and the discriminant is $ \Delta := b_B^2 B - 4\delta \omega(B - \delta)(B - \sigma )$.

Now, suppose $\cB\st \equiv B(x\st)$, where $B:\mathbb{X} \rightarrow (\sigma, \infty)$ is a continuously differentiable function, and 
let $\mathbb{Y} := (0,1]\times[0,1]$ and $y^*(x) := (I_{B(x)}^*, R_{B(x)}^*)$.
For a fixed $x\st \equiv x$, the following is a strict Lyapunov function for \eqref{eq:SIRS} on $\mathbb{Y}$:
\begin{align}
    {\mathcal{U}}(I,R;x) :=& \; {\tilde{\mathcal{U}}}(I,R;B(x)), 
        \label{eq:Lyapunov_SIRS}
\end{align}
where
\begin{align*}
    {\widetilde{\mathcal{U}}}(I,R;B) :=& (I- I_B^*)+I_B^*\ln{\frac{I_B^*}{I}}+\frac{a_B}{2}(R-R_B^*)^2, \nonumber
\end{align*}
and $a_B:= B /(\gamma +\delta R_B^*)$. The derivative of \eqref{eq:Lyapunov_SIRS} along trajectories is 
\begin{align*}
    \frac{d}{dt}{\widetilde{\mathcal{U}}}(I\st,R\st;B) 
=& \ -(B - \delta)(I\st - I_B^*)^2 \\ 
& - a_B(\omega - \delta I\st)(R\st - R_B^*)^2, 
\end{align*}
and it is negative for any $(I\st,R\st) \in \mathbb{Y}\setminus \{ y^*(x) \}$.

Since \eqref{eq:Lyapunov_SIRS} is a strict Lyapunov function when $x\st$ is constant, it satisfies conditions (i) and (iii). As $\mathcal{U}$ is a continuous function, its sublevel sets are closed. Moreover, because the sublevel sets are also contained in $\mathbb{Y}\times\mathbb{X}$, which is a bounded set, they are also bounded and, hence, condition (ii) holds. Lastly, we can verify condition (iv) as follows.
\begin{align*}
    \frac{\partial \mathcal{U}}{\partial x}(I_{B(x)}^*,R_{B(x)}^*;x) 
    & = \frac{\partial {\widetilde{\mathcal{U}}}}{\partial B}(I_{B(x)}^*,R_{B(x)}^*;B(x)) \frac{\partial B}{\partial x}(x)
    = 0,  %
\end{align*}
where the second equality follows from $\frac{\partial {\widetilde{\mathcal{U}}}}{\partial B}(I_{B(x)}^*,R_{B(x)}^*;B(x)) = 0$.
\subsection{Designing an Intervention Policy for Epidemics with Nonlinear Infection Rate}
\label{ssec:application}

We consider a modification of the EPG studied in \cite{certorio_epidemic_2022}, which was described in \S \ref*{ssec:epg}, as an application of Theorem \ref{thm:convergence} to a dynamic payoff design problem. We aim to mitigate an epidemic outbreak and reduce the endemic level of infected agents while guaranteeing that the long-term cost of the policy maker does not exceed some available budget $c^*$.

 The study in \cite{certorio_epidemic_2022} considers the average transmission rate that depends linearly on the population state, with $\cB\st \equiv {\beta}^\top x\st $, where ${\beta}\in \mathbb{R}_{>0}^n$. Such dependency on $x\st$ is consistent with the choices of the susceptible agents determining the likelihood of contracting the disease when exposed, e.g., choosing to wear masks or getting vaccinated. In their model, the proportion of susceptible agents following the $i$-th strategy at time $t$ is $x_i\st S\st$, and for those agents the rate of new infections is equal to $\beta_i x_i\st S\st I\st$.

Suppose that we allow the average transmission rate to depend on both the choices of the susceptible agents and those of the infected agents. For example, an infected agent that takes no preventive measures is more likely to transmit the disease than another infected agent that does take preventive measures. In this case, the rate of new infections among susceptible agents following strategy $i$ due to the exposure to infected agents adopting strategy $j$ would be $\beta_{ij} x_i\st S\st x_j\st I\st$. Therefore, the average transmission rate is given by 
\begin{align*}
    \cB\st \equiv B(x\st) := x\st^\top  Q x\st, \quad t \geq 0,
\end{align*}
with $Q\in \mathbb{R}^{n \times n}$ being a positive matrix with the elements $Q_{ij} = \beta_{ij}$.
We assume that the disease is too infectious to be eradicated so that $B(x) \geq \sigma$ for all $x\in \mathbb{X}$.

We aim to select a target population equilibrium $x^*$ that minimizes the transmission rate subject to a budget constraint. In order to find $x^*$ we solve
\begin{align*}
    x^* \in \argmin_{z \in \mathbb{X}} B(z) \ \text{ s.t. } c^\top z - \min_i c_i\leq c^*,
\end{align*}
where $c$ is the vector of intrinsic costs of the strategies, and $c^*$ is the long-term budget available to the policy maker. 

After determining $x^*$, we select $k_1, k_2, k_3 > 0 $ and use the payoff mechanism described in Theorem~\ref{thm:convergence} to lead the population and epidemic states to the selected equilibrium.

\begin{example} \label{example1} 
Consider a disease with parameters: $\delta = 0.005$, $\zeta=0$, $\theta=0.0002$, $\gamma=0.1$ (mean recovery period $\sim$ 10 days) and $\bar{\omega}=0.011$~(mean immunity period $\sim$ 91 days). Agents have three available strategies with
\begin{align*}
    Q = \begin{bmatrix}
        0.13 & 0.18 & 0.2 \\
        0.16 & 0.22 & 0.23 \\
        0.17 & 0.28 & 0.5
    \end{bmatrix}, \ \mbox{ and } \
        c &= \begin{bmatrix}
        0.2 \\  0.1 \\ 0
    \end{bmatrix}.
\end{align*}

The initial conditions are $I\sta{0} = 0.019$ and $R\sta{0} = 0.172$ for the epidemic, $x\sta{0} = (1\; 0\; 0)^\top$ for the EDM, and $q\sta{0}=(0\; 0\; 0)^\top$ for the payoff dynamics. The long-term budget of the policy maker is $c^*=0.1$, and we select $p^*=\mathbf{0}$, which yields $x^* \approx (1\; 10\; 1)/12$ and $y^*(x^*)\approx (5.1 \%,46.9 \%) $. Our goal is to design $G$ and $H$ so that ${I\st} \leq 10\%$ for all $t \geq 0$.
\end{example}

For simplicity, we select $q_0=-c$ and $p^*=p(0)=\mathbf{0}$ and $k_3=1$, as $k_3$ does not affect the bound on $I\st$ if $p^*=\mathbf{0}$. 
Based on \eqref{ineq:L_less_L0}, we look for values of $k_1$ and $k_2$ that meet our requirement that ${I\st} \leq 10\%$ for all $t \geq 0$,  by solving %
\begin{align}
    I_\text{max}(k_1,k_2) := \max_{I,R,x} \quad &I 
        \nonumber \\
    \text{s.t.} ~ x&\in\mathbb{X}, \nonumber \\ 
                I,R&\geq0,       \nonumber \\
                I+R&\leq1,       \nonumber \\
        \mathcal{L}(y,x,0)&\leq \mathcal{L}(y\sta{0},x\sta{0},0), \label{eq:I_max_L_dependency}
\end{align}
where $k_1$ and $k_2$ affect constraint \eqref{eq:I_max_L_dependency}.

\begin{figure}
    \centering
    \includegraphics[width=0.9\columnwidth]{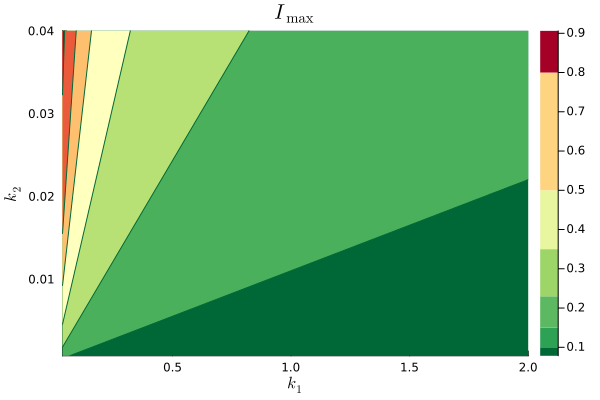}
    \caption{Bound $I_\text{max}(k_1,k_2)$, for the conditions in Example \ref{example1}, when varying $k_1$ and  $k_2$.}
    \label{fig:I_payoff}
\end{figure}

We solve numerically the optimization above for several values of $k_1$ and $k_2$, and show the results in Fig. \ref{fig:I_payoff}. The requirement that ${I\st} \leq 10\%$ is met for $k_1,k_2$ in the region on the bottom right of the plot, and we select  $k_1=2$ and $k_2=0.022$. We then use the reward mechanism given by Theorem \ref{thm:convergence} to guide the population to the desired equilibrium.

\begin{figure}
    \centering
    \includegraphics[width=\columnwidth]{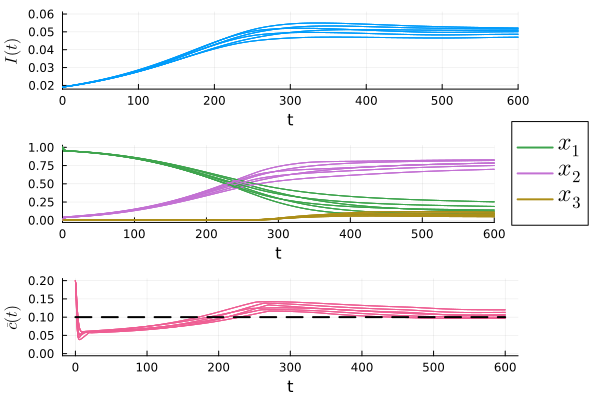}
    \caption{Simulation of Example \ref{example1}, for many different learning rules, using $k_1=2,k_2=0.022,k_3=1$.}
    \label{fig:many_learning_rules_well}
\end{figure}

\begin{figure}
    \centering
    \includegraphics[width=\columnwidth]{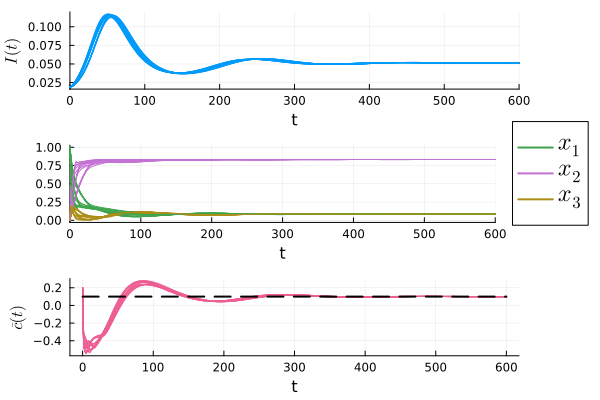}
    \caption{Simulation of Example \ref{example1}, for many different learning rules, using $k_1=k_2=k_3=1$.}
    \label{fig:many_learning_rules_poorly}
\end{figure}

The simulation results for several different learning rules that satisfy condition (iv) in Theorem~\ref{thm:convergence} are shown in Fig.~\ref{fig:many_learning_rules_well}.\footnote{See the simulation code at \href{https://github.com/jcert/incentive-design-coupled-dynamics}{github/jcert/incentive-design-coupled-dynamics} for more details on the learning rules that were used.} It is clear that they all converge and satisfy the requirement that ${I\st} \leq 10\%$ for all $t \geq 0$. Had we not used the bound, $I_\text{max}(k_1,k_2)$, to determine the parameters of the reward mechanism, the bound on the peak of infections may have been violated, as shown in Fig. \ref{fig:many_learning_rules_poorly}. In both figures we observe that the instantaneous cost, $\Bar{c}\st$, converges to $c^*$, which is represented as a dashed black line in the plots.

\section{Conclusion}
\label{sec:conclusion}

We studied a large population of learning agents whose strategic choices influence the dynamics of an ES we seek to stabilize at a desired equilibrium. Our framework can be used to design a dynamic payoff mechanism that guarantees the convergence of both the population and the ES (Theorem \ref{thm:convergence}). When the conditions on the ES stated in the theorem are met, the designed incentives can stabilize more general systems than previously considered.

We also presented example systems that satisfy the conditions of our main result (\S \ref{ssec:Korobeinikov} and \S \ref{ssec:epg}) and applied our framework to design incentives that mitigate an epidemic with nonlinear infection rates subject to long-time budget constraints (\S \ref{ssec:application}). Unlike the incentives designed in the previous studies of EPG, our payoff mechanism is guaranteed to have a bound on the instantaneous reward offered to the population.  

In future research we plan to extend the results by relaxing the assumptions on the ES to be only local and to allow the EDM to depend on the ES states. Another direction we are interested in pursuing is to examine design problems where the payoff of certain strategies can only be partially designed.

\addtolength{\textheight}{-17cm}   %

\bibliographystyle{IEEEtran}
\bibliography{ref}

\end{document}